# Muon Ring and FCC-*ee* / CEPC Based Antimuon-Electron Colliders

Dilara Akturk[1*], Burak Dagli[1], Saleh Sultansoy[1,2]

[1]TOBB University of Economics and Technology, Ankara, Türkiye

[2]ANAS Institute of Physics, Baku, Azerbaijan

**Abstract**

Recently, the construction of an antimuon-electron collider, $\mu$TRISTAN, at KEK has been proposed. We argue that the construction of a similar muon ring tangential to FCC-*ee* and CEPC will give an opportunity to realize antimuon-electron collisions at higher center-of-mass energies. Moreover, the same ring may be used later to realize energy-frontier antimuon-proton colliders based on FCC-*pp* and SppC. Similarly, changing of electron ring in $\mu$TRISTAN project into proton ring will give opportunity to investigate lepton-hadron collisions at ~2 TeV center-of-mass energies. In this paper the main parameters of proposed colliders have been studied. It is shown that sufficiently high luminosities can be achieved for all proposals under consideration.

## 1. Introduction

Detailed investigation of properties of the Higgs boson, discovered in 2012 at the LHC [1, 2] is among the most important aims of high energy physics. In this context, lepton colliders will play a crucial role due to the clean experimental environment. For this reason, linear electron-positron colliders (ILC [3] and CLIC [4]), muon colliders [5] and circular electron-positron colliders (FCC-*ee* [6], CEPC [7]) are being developed.

Muon-electron colliders, proposed 30 years ago in [8-10], can also make a significant contribution to the study of the properties of the Higgs boson. A few papers were subsequently published regarding the physics research potential of these colliders (see for example [11-16]). Certainly, muon-electron colliders have significant potential for the investigation of lepton flavor violation and flavor-changing neutral current phenomena.

The first comprehensive paper [17] on the parameters of a $\mu e$ collider has been published recently. Herein, the collision of 30 GeV electrons and 1 TeV antimuons in the 3 km tunnel in KEK is proposed. 30 GeV electron beam had already been obtained in TRISTAN. It is planned to use the method developed at JPARC to obtain a low emittance $\mu^+$ beam [18] which then will be accelerated up to TeV energy.

In this paper, we propose to use FCC-*ee* and CEPC instead of TRISTAN for the electron beam. The proposed colliders have two important advantages:

• Higher $\mu e$ center-of-mass energies,

• Since FCC-*ee* and CEPC are planned to be converted into proton-proton colliders in the future, the same muon ring will enable the establishment of energy-frontier antimuon-proton colliders.

We use AloHEP (A luminosity optimizer for High Energy Physics) software [19-21] to calculate center-of-mass energy ($\sqrt{S}$), luminosity (L), transverse beam sizes ($\sigma_x$, $\sigma_y$) and beam-beam tune-shift ($\xi_x$, $\xi_y$) parameters. This software was developed several years ago for estimation of main parameters of linac-ring type *ep* colliders. Later, AloHEP was upgraded for all types of colliders (linear, circular, and linac-ring) and colliding beams (electron, positron, muon, proton, and nuclei).

In the second section, the parameters of the *μ*TRISTAN proposal were compared with AloHEP software predictions, in addition, tune-shift values were calculated. The parameters of *μe* colliders based on FCC-*ee* and CEPC are considered in the third and fourth sections, respectively. In the fifth section, the parameters of multi-TeV scale antimuon-proton colliders that can be installed at the next stage are calculated. In the last section, our conclusions and suggestions are given.

## 2. *μ*TRISTAN

In Table 1, we present electron and antimuon beam parameters of the *μ*TRISTAN proposal [17].

**Table 1**. Main parameters of *μ*TRISTAN's *μe* option

| Parameter | Antimuon | Electron |
|---|---|---|
| Number of Particle per Bunch [$10^{10}$] | 1.4 | 6.2 |
| Beam Energy [GeV] | 1000 | 30 |
| Horizontal β Function @ IP [cm] | 3 | 3 |
| Vertical β Function @ IP [cm] | 0.7 | 0.7 |
| Bunch Length [mm] | 2 | 2 |
| Norm. Horizontal Emittance [μm] | 4 | 4 |
| Norm. Vertical Emittance [μm] | 4 | 4 |
| Number of Bunches per Ring | 40 | 40 |
| Collision Frequency [MHz] | 4 | 4 |
| Circumference [km] | 3 | 3 |

By entering the values from Table 1 into AloHEP, the parameters of the *μe* collider shown in Table 2 are obtained (parameters from the original article [17] are given in parentheses). It is seen that AloHEP outputs are in full agreement with the corresponding values from [17].

**Table 2.** Main parameters of the *μ*TRISTAN's *μe* collider calculated by AloHEP

| Parameter | |
|---|---|
| $\sqrt{S}$ [GeV] | 346 (346) |
| L [$10^{33}$ cm$^{-2}$s$^{-1}$] | 4.54 (4.6) |
| $\sigma_x$ [μm] | 3.55 (3.6) |
| $\sigma_y$ [μm] | 1.72 (1.7) |

Let us mention that beam-beam tune-shift values, which should be maintained under control in collider designs, were not considered in the original article. The values obtained using AloHEP software are given in Table 3.

Table 3. Beam-beam tune-shift values of the μTRISTAN's μe collider

| Parameter | Muon | Electron |
|---|---|---|
| Tuneshift ($\xi_x$) | $1.10\times10^{-2}$ | $8.29\times10^{-2}$ |
| Tuneshift ($\xi_y$) | $2.28\times10^{-2}$ | $1.71\times10^{-1}$ |

As can be seen, while muon beam tune-shifts are sufficiently low, the tune-shift values of the electron beam are close to the limit values.

## 3. μFCC-*ee*

This section is devoted to a possible adaptation of the μTRISTAN proposal for FCC-*ee*. FCC-*ee* [6] is designed in three stages with the center-of-mass energies of 92 GeV (Z resonance region), 240 GeV (ZH, WW, ZZ region) and 366 GeV ($t\bar{t}$ region). The electron beam parameters designed for these stages are given in Table 4 (parameters given in the PDG [22] are used).

Table 4. Main parameters of the FCC-*ee* collider

| Parameter | Stage 1 | Stage 2 | Stage 3 |
|---|---|---|---|
| Number of Particle per Bunch [$10^{11}$] | 2.1 | 1.2 | 1.6 |
| Particle Beam Energy [GeV] | 46 | 120 | 183 |
| Horizontal β Function @ IP [cm] | 11 | 24 | 100 |
| Vertical β Function @ IP [cm] | 0.07 | 0.1 | 0.16 |
| Bunch Length [mm] | 15.5 | 4.7 | 2.2 |
| Norm. Horizontal Emittance [μm] | 63 | 167 | 568 |
| Norm. Vertical Emittance [μm] | 0.17 | 0.3 | 0.6 |
| Collision Frequency [MHz] | 37 | 1.5 | 0.2 |
| Circumference [km] | 90.66 | 90.66 | 90.66 |

By entering the parameters of antimuon beam from Table 1 and electron beams from Table 4 into AloHEP, we obtained the results for main parameters of μFCC-*ee* colliders which are given in Table 5.

Table 5. Main parameters of μFCC-*ee* colliders

| Parameter | Stage 1 | | Stage 2 | | Stage 3 | |
|---|---|---|---|---|---|---|
| $\sqrt{S}$ [GeV] | 429 | | 693 | | 856 | |
| L [$10^{33}$ cm$^{-2}$s$^{-1}$] | 6.20 | | 0.89 | | 0.05 | |
| Parameter | Muon | Electron | Muon | Electron | Muon | Electron |
| $\sigma_x$ [μm] | 8.8 | 8.8 | 13 | 13 | 40 | 40 |
| $\sigma_y$ [μm] | 1.7 | 1.7 | 1.7 | 1.7 | 1.7 | 1.7 |
| Tuneshift ($\xi_x$) | $1.9\times10^{-2}$ | $3.1\times10^{-1}$ | $7.6\times10^{-3}$ | $1.0\times10^{-1}$ | $3.6\times10^{-3}$ | $1.4\times10^{-2}$ |
| Tuneshift ($\xi_y$) | $9.6\times10^{-2}$ | 1.6 | $5.8\times10^{-2}$ | $7.8\times10^{-1}$ | $8.4\times10^{-2}$ | $3.3\times10^{-1}$ |

As can be seen, although there is no problem with the muon tune-shift, values of the tune-shift of the electron beams are unacceptably high. It is necessary to reduce the electron tune-shift to around 0.1. For example, in the case of FCC-*ee* stage 1, this can be achieved by reducing the number of muons in the bunch by a factor of 10. As a result, the luminosity will decrease by 10 times. To compensate for this decrease in luminosity, the number of electrons in the bunch can be increased by an order with a corresponding decrease in the number of bunches in the electron ring in order to keep synchrotron radiation power under control (unfortunately this cannot be applied for stages 2 and 3). Another option is the increase of the emittance of the electron beam together with the proportional decrease of the beta function. Certainly, a combination of these two methods will allow solving the tune-shift problem.

## 4. μCEPC

CEPC [7] is also designed in three stages with the center-of-mass energies of 92 GeV, 240 GeV and 360 GeV. The electron beam parameters designed for these stages are given in Table 6 (parameters given in the PDG [22] are used).

Table 6. Main parameters of the CEPC proposal

| Parameter | CEPC 1 | CEPC 2 | CEPC 3 |
|---|---|---|---|
| Number of Particle per Bunch [$10^{11}$] | 1.4 | 1.3 | 2 |
| Beam Energy [GeV] | 46 | 120 | 180 |
| Horizontal β Function @ IP [cm] | 13 | 33 | 104 |
| Vertical β Function @ IP [cm] | 0.09 | 0.1 | 0.27 |
| Bunch Length [mm] | 8.7 | 3.9 | 2.9 |
| Norm. Horizontal Emittance [μm] | 24 | 150 | 493 |
| Norm. Vertical Emittance [μm] | 0.12 | 0.3 | 1.7 |
| Collision Frequency [MHz] | 36 | 0.75 | 0.11 |
| Circumference [km] | 100 | 100 | 100 |

The main parameters of μCEPC colliders obtained by entering the parameters from Tables 1 and 6 into AloHEP are given in Table 7.

Table 7. Main parameters of μCEPC colliders

| Parameter | CEPC 1 | | CEPC 2 | | CEPC 3 | |
|---|---|---|---|---|---|---|
| √S [GeV] | 429 | | 693 | | 848 | |
| L [$10^{33}$ cm$^{-2}$s$^{-1}$] | 6.18 | | 0.43 | | 0.04 | |
| Parameter | Muon | Electron | Muon | Electron | Muon | Electron |
| $\sigma_x$ [μm] | 5.9 | 5.9 | 15 | 15 | 38 | 38 |
| $\sigma_y$ [μm] | 1.7 | 1.7 | 1.7 | 1.7 | 1.7 | 1.7 |
| Tuneshift ($\xi_x$) | $1.7\times10^{-2}$ | $8.3\times10^{-1}$ | $7.5\times10^{-3}$ | $9.9\times10^{-2}$ | $4.7\times10^{-3}$ | $9.3\times10^{-3}$ |
| Tuneshift ($\xi_y$) | $5.9\times10^{-2}$ | 2.9 | $6.3\times10^{-2}$ | $8.4\times10^{-1}$ | $1.0\times10^{-1}$ | $2.1\times10^{-1}$ |

Again, values of the tune-shift of the electron beams are unacceptably high. This problem can be solved in the same manner as suggested for μFCC-ee in the previous section.

## 5. Next stage: Antimuon-proton colliders

As we mentioned in the introduction, FCC-ee and CEPC will be converted into hadron colliders in the next stage. The antimuon beam to be installed for the μe collider will allow high-energy $\mu p$ collisions to be achieved by colliding it with the proton beams of FCC-*hh* and SppC. $\mu$TRISTAN will also enable the exploration of high-energy antimuon-proton collisions if the electron beam of TRISTAN is replaced with a proton beam in the next stage.

**5.1. FCC based *μp* collider**

FCC-based $\mu p$ colliders have been proposed in [23] (for a review of muon-proton collider proposals, see [24]). Here, the parameters of the FCC-based $\mu p$ collider are re-estimated using the $\mu^+$ beam parameters of $\mu$TRISTAN. The parameters of the FCC proton beam [25] are given in Table 8.

**Table 8.** FCC proton beam parameters (nominal and ERL60 upgraded)

| Parameter | Nominal | Upgraded |
|---|---|---|
| Number of Particle per Bunch [$10^{11}$] | 1.0 | 1.0 |
| Beam Energy [TeV] | 50 | 50 |
| Horizontal β Function @ IP [m] | 1.1 | 0.15 |
| Vertical β Function @ IP [m] | 1.1 | 0.15 |
| Norm. Horizontal Emittance [μm] | 2.2 | 2.2 |
| Norm. Vertical Emittance [μm] | 2.2 | 2.2 |
| Revolution Frequency [Hz] | 2998 | 2998 |
| Number of Bunches per Ring | 10400 | 10400 |
| Circumference [km] | 100 | 100 |

By entering the parameters from Tables 1 and 8 into AloHEP, we obtained the results for the main parameters of $\mu$FCC colliders which are given in Table 9. As can be seen from the last two rows of the table, there is no problem with the tune-shift values.

**Table 9.** Main parameters of the FCC based $\mu p$ collider

| Parameter | FCC Nominal | | FCC Upgraded | |
|---|---|---|---|---|
| $\sqrt{S}$ [TeV] | 14.14 | | 14.14 | |
| L [$10^{32}$ cm$^{-2}$s$^{-1}$] | 9.8 | | 50 | |
| Parameter | Proton | Muon | Proton | Muon |
| $\sigma_x$ [μm] | 6.7 | 6.7 | 3.6 | 3.6 |
| $\sigma_y$ [μm] | 6.7 | 6.7 | 2.5 | 2.5 |
| Tuneshift ($\xi_x$) | $7.8\times10^{-4}$ | $2.7\times10^{-2}$ | $6.4\times10^{-4}$ | $2.2\times10^{-2}$ |
| Tuneshift ($\xi_y$) | $7.8\times10^{-4}$ | $2.7\times10^{-2}$ | $9.2\times10^{-4}$ | $3.2\times10^{-2}$ |

Let us emphasize that center-of-mass energy value is 4 times higher than that of ERL60 and FCC based *ep* collider [25]. Since the luminosity values are of the same order, the physics research potential of the $\mu p$ collider will be much higher than that of the *ep* collider.

**5.2. SppC based *μp* collider**

SppC based $\mu p$ colliders have been proposed in [26]. Herein, the parameters of the SppC based $\mu p$ collider will be determined using the antimuon beam parameters of $\mu$TRISTAN. The parameters of the SppC proton beam are given in Table 10.

Table 10. SppC proton beam parameters

| Parameter | SppC 1 | SppC 2 |
|---|---|---|
| Number of Particle per Bunch [$10^{11}$] | 2.0 | 2.0 |
| Beam Energy [TeV] | 35.6 | 68 |
| Horizontal β Function @ IP [cm] | 75 | 24 |
| Vertical β Function @ IP [cm] | 75 | 24 |
| Norm. Horizontal Emittance [μm] | 4.1 | 3.05 |
| Norm. Vertical Emittance [μm] | 4.1 | 3.05 |
| Revolution Frequency [Hz] | 5485 | 3000 |
| Number of Bunches per Ring | 5835 | 10667 |
| Circumference [km] | 54.7 | 100 |

The main parameters of *μ*SppC colliders obtained using AloHEP are given in Table 11.

Table 11. Main parameters of the SppC based *μp* colliders

| Parameter | SppC 1 | | SppC 2 | |
|---|---|---|---|---|
| $\sqrt{S}$ [TeV] | 11.93 | | 16.49 | |
| L [$10^{32}$ cm$^{-2}$s$^{-1}$] | 6.9 | | 79 | |
| Parameter | Proton | Muon | Proton | Muon |
| σ$_x$ [μm] | 9.0 | 9.0 | 3.6 | 3.6 |
| σ$_y$ [μm] | 9.0 | 9.0 | 3.2 | 3.2 |
| Tuneshift (ξ$_x$) | 4.2×10$^{-4}$ | 5.5×10$^{-2}$ | 5.3×10$^{-4}$ | 5.2×10$^{-2}$ |
| Tuneshift (ξ$_y$) | 4.2×10$^{-4}$ | 5.5×10$^{-2}$ | 5.9×10$^{-4}$ | 5.8×10$^{-2}$ |

Certainly, *μ*SppC has a huge potential for both the SM (especially QCD basics) and BSM (especially second family related) physics searches, similar to that of *μ*FCC.

### 5.3. *μ*TRISTAN based *μp* collider

In principle, *μ*TRISTAN can also be converted into the *μp* collider: the TRISTAN ring with 3 km length can be used for a proton accelerator at a later stage. Proton beams with 0.85 TeV energy can be obtained with 8 Tesla and 1.7 TeV energy with 16 Tesla bending magnets. The parameters of the proton beam are presented in Table 12 (for proton bunch population, beta functions at IP and emittance parameters values given in last column of Table 8 are used).

Table 12. Parameters of the TRISTAN proton beam

| Parameter | 8/16 Tesla |
|---|---|
| Number of Particle per Bunch [$10^{11}$] | 1.0 |
| Particle Beam Energy [TeV] | 0.85/1.7 |
| Horizontal β Function @ IP [m] | 0.15 |
| Vertical β Function @ IP [m] | 0.15 |
| Norm. Horizontal Emittance [μm] | 2.2 |
| Norm. Vertical Emittance [μm] | 2.2 |
| Revolution Frequency [MHz] | 0.1 |
| Number of Bunches per Ring | 20 |
| Circumference [km] | 3.0 |

By entering the parameters from Tables 1 and 12 into AloHEP, we obtained the results for the main parameters of $\mu$TRISTAN based $\mu p$ colliders which are given in Table 13.

**Table 13.** Main parameters of the $\mu$TRISTAN based $\mu p$ colliders

| Parameter | 8 Tesla Magnets | | 16 Tesla Magnets | |
|---|---|---|---|---|
| $\sqrt{s}$ [TeV] | 1.84 | | 2.61 | |
| L [$10^{32}$ cm$^{-2}$s$^{-1}$] | 1.22 | | 2.45 | |
| Parameter | Proton | Muon | Proton | Muon |
| $\sigma_x$ [μm] | 19.1 | 19.1 | 13.5 | 13.5 |
| $\sigma_y$ [μm] | 19.1 | 19.1 | 13.5 | 13.5 |
| Tuneshift ($\xi_x$) | $7.8\times10^{-4}$ | $2.7\times10^{-2}$ | $7.8\times10^{-4}$ | $2.7\times10^{-2}$ |
| Tuneshift ($\xi_y$) | $7.8\times10^{-4}$ | $2.7\times10^{-2}$ | $7.8\times10^{-4}$ | $2.7\times10^{-2}$ |

Let us mention that center-of-mass energy values are essentially higher than that of the LHeC [27], while the luminosity values are one order magnitude lower. Concerning clarification of QCD basics colliders under consideration will be much more powerful instruments than the LHeC.

## 6. Conclusion

The main advantage of $\mu$TRISTAN is that it can be realized earlier (e.g. in the 2030s), while $\mu$FCC-$ee$ and $\mu$CEPC can be implemented in the 2040s at the earliest. But the last two options have a different and very important advantage: FCC-$ee$ and CEPC will later be converted into hadron colliders, so that the use of the muon ring at this stage will allow the construction of multi-TeV scale antimuon-hadron colliders. In principle, $\mu$TRISTAN can also be converted into antimuon-proton collider, if electron ring will be replaced with proton ring.

**Table 14.** Main parameters of $\mu e$ and $\mu p$ colliders

| Colliders | | | $\sqrt{s}$ | L (cm$^{-2}$s$^{-1}$) |
|---|---|---|---|---|
| $\mu e$ | $\mu$TRISTAN | | 346 GeV | $4.5\times10^{33}$ |
| | $\mu$FCC-$ee$ | Stage 1 | 429 GeV | $6.2\times10^{33}$ |
| | | Stage 2 | 693 GeV | $8.9\times10^{32}$ |
| | | Stage 3 | 856 GeV | $5.0\times10^{31}$ |
| | $\mu$CEPC | Stage 1 | 429 GeV | $6.2\times10^{33}$ |
| | | Stage 2 | 693 GeV | $4.3\times10^{32}$ |
| | | Stage 3 | 848 GeV | $4.0\times10^{31}$ |
| $\mu p$ | $\mu$FCC | | 14.14 TeV | $5.0\times10^{33}$ |
| | $\mu$SppC | Option 1 | 11.93 TeV | $6.9\times10^{32}$ |
| | | Option 2 | 16.49 TeV | $7.9\times10^{33}$ |
| | $\mu$TRISTAN | 8 Tesla | 1.84 TeV | $1.2\times10^{32}$ |
| | | 16 Tesla | 2.61 TeV | $2.5\times10^{32}$ |

In Table 14, we present center-of-mass energies and luminosities of μTRISTAN as well as μe and μp colliders proposed in this paper. It is seen that luminosities of μTRISTAN, μFCC-ee stage 1, μCEPC stage 1, μFCC and μSppC option 2 well exceed $10^{33}$ cm$^{-2}$s$^{-1}$; luminosities of μFCC-ee stage 2, μCEPC stage 2, μSppC option 1 and μTRISTAN based μp collider lie between $10^{32}$ cm$^{-2}$s$^{-1}$ and $10^{33}$ cm$^{-2}$s$^{-1}$; luminosities of μFCC-ee stage 3 and μCEPC stage 3 is below $10^{32}$ cm$^{-2}$s$^{-1}$. Concerning Higgs boson production at μe colliders, only μTRISTAN, μFCC-ee stage 1, and μCEPC stage 1 can make an important contribution, since the increase of energy at stages 2 and 3 does not compensate essential decreasing of luminosities. On the other hand, an increase in energy may be important for lepton flavor violation and flavor-changing neutral current phenomena.

Finally, let us emphasize that the antimuon-electron and antimuon-proton colliders under consideration can be realized before the muon colliders. The reason is that while $\mu^-$ beams with emittance, which is sufficiently small for the construction of muon colliders, have not yet been achieved, there is an established technology to create a low emittance $\mu^+$ beam by using ultra-cold muons [18].

**Acknowledgement:**

Authors are grateful to Arif Ozturk for useful discussions.

**References**

[1] G. Aad et al., "Observation of a new particle in the search for the Standard Model Higgs boson with the ATLAS detector at the LHC", Physics Letters B 716.1, 1 (2012)

[2] S. Chatrchyan et al., "Observation of a new boson at a mass of 125 GeV with the CMS experiment at the LHC", Physics Letters B 716.1, 30 (2012)

[3] T. Behnke et al., "The international linear collider technical design report-volume 1: Executive summary", arXiv:1306.6327 (2013)

[4] T. K. Charles et al., "The Compact Linear Collider (CLIC) - 2018 Summary Report", arXiv:1812.06018 (2018)

[5] J. P. Delahaye et al., "Muon colliders", arXiv:1901.06150 (2019)

[6] A. E. A. Abada, et al., "FCC-ee: the lepton collider", The European Physical Journal Special Topics, 228, 261-623 (2019)

[7] CEPC Study Group, "CEPC conceptual design report: Volume 1-accelerator", arXiv:1809.00285 (2018)

[8] G. W.-S. Hou, "Possible resonances in μ$^+$e$^-$ → μ$^-$e$^+$ collisions," Nuclear Physics B - Proceedings Supplements, vol. 51, no. 1, pp. 40–49 (1996)

[9] S. Y. Choi, C. S. Kim, Y. J. Kwon, and S. H. Lee, "High energy FCNC search through eμ colliders," Physical Review D, vol. 57, no. 11, pp. 7023–7026 (1998)

[10] V. Barger, S. Pakvasa, and X. Tata, "Are eμ colliders interesting?," Physics Letters B, vol. 415, no. 2, pp. 200–204 (1997)

[11] S.Y. Choi, C.S. Kim, Y.J. Kwon and S. H. Lee, "High Energy FCNC search through eμ Colliders", Physical Review D, 57 7023-7026 (1998)

[12] J. K. Singhal, S. Singh and A. K. Nagawat, "Possible exotic neutrino signature in electron muon collisions", arXiv:hep-ph/0703136 (2007)

[13] F. Bossi and P. Ciafaloni, "Lepton Flavor Violation at muon-electron colliders", JHEP 10, 033 (2020)

[14] A. O. Bouzas and F. Larios, "Two-to-Two Processes at an Electron-Muon Collider", Adv. High Energy Phys. 2022, 3603613 (2022)

[15] M. Lu et al., "The physics case for an electron-muon collider", Adv. High Energy Phys. 2021 6693618 (2021)

[16] P. S. Bhupal, J. Heeck and A. Thapa, "Neutrino mass models at μTRISTAN", The European Physical Journal C, 84.2 148 (2024)

[17] Y. Hamada et al., "$\mu$TRISTAN", Progress of Theoretical and Experimental Physics, 2022, 5, 053B02 (2022)

[18] Y. Kondo et al., "Re-acceleration of ultra cold muon in J-PARC muon facility", in Proc. 9th Int. Particle Accelerator Conference, p. 6 (2018)

[19] B. Dagli, S. Sultansoy, B. Ketenoglu and B. B. Oner, "Beam-Beam Simulations for Lepton-Hadron Colliders: AloHEP Software", 12th Int. Particle Acc. Conf (IPAC'21)

[20] B. Dagli and B. B. Oner, "A Luminosity Optimizer for High Energy Physics", TOBB ETU High Energy Physics http://yef.etu.edu.tr/ALOHEP2_eng.html (2022)

[21] B. Dagli and A. Ozturk, "ALOHEP (A Luminosity Optimizer for High Energy Physics)", https://github.com/yefetu/ALOHEP (2022)

[22] V. Shiltsev and F. Zimmermann, "Accelerator Physics of Colliders", in Particle Data Group, Progress of Theoretical and Experimental Physics, Volume 2022, Issue 8 (2022)

[23] Y. C. Acar et al., "Future circular collider based lepton–hadron and photon–hadron colliders: Luminosity and physics", Nuclear Instruments and Methods in Physics Research Section A 871 47–53 (2017)

[24] B. Ketenoglu, B. Dagli, A. Ozturk and S. Sultansoy, "Review of muon-proton and muon-nucleus collider proposals", Modern Physics Letters A, 37(37n38), 2230013 (2022)

[25] A. Abada et al., "FCC-hh: The Hadron Collider", The European Physical Journal Special Topics 228, 755–1107 (2019)

[26] A. C. Canbay et al., "SppC based energy frontier lepton-proton colliders: Luminosity and physics", Advances in High Energy Physics Volume 2017, 4021493 (2017)

[27] P. Agostini, et al., "The large hadron–electron collider at the HL-LHC", Journal of Physics G: Nuclear and Particle Physics, 48(11), 110501 (2021)